\RequirePackage{fix-cm}
\documentclass[twocolumn]{svjour3}          % twocolumn
\smartqed  % flush right qed marks, e.g. at end of proof
\usepackage{graphicx}
\newtheorem{thm}{Theorem}[section]

\newtheorem{lem}[thm]{Lemma}

\newtheorem{defn}[thm]{Definition}

\usepackage{amsmath}
\usepackage{bm}
 
\journalname{Journal of Signal Processing Systems}

\begin{document}

\title{Stability of the Stochastic Gradient Method for an Approximated Large Scale Kernel Machine
}


\author{Aven Samareh        \and
        Mahshid Salemi Parizi %etc.
}


\institute{Aven Samareh \at
              Department of Industrial and Systems Engineering \\
              University of Washington, Seattle, WA 98105, USA\\
              \email{asamareh@uw.edu}  \\
              ORCID:0000-0002-2630-8041%  \\
}

\date{Received: date / Accepted: date}

\maketitle

\begin{abstract}
In this paper we measured the stability of stochastic gradient method (SGM) for learning an approximated Fourier primal support vector machine. The stability of an algorithm is considered by measuring the generalization error in terms of the absolute difference between the test and the training error. Our problem is to learn an approximated kernel function using random Fourier features for a binary classification problem via online convex optimization settings. For a convex, Lipschitz continuous and smooth loss function, given reasonable number of iterations stochastic gradient method is stable. We showed that with a high probability SGM generalizes well for an approximated kernel under given assumptions. We empirically verified the theoretical findings for different parameters using several data sets.


\end{abstract}

\section{Introduction}\label{sec:introduction}
The stochastic gradient method (SGM) is widely used as an optimization tool in many machine learning applications including (linear) support vector machines \cite{shalev2011pegasos,zhu2009p}, logistic regression \cite{zhuang2015distributed,bottou2012stochastic}, graphical models \cite{dean2012large,poon2011sum,bonnabel2013stochastic} and deep learning \cite{deng2013recent,krizhevsky2012imagenet,haykin2009neural}. SGM computes the estimates of the gradient on the basis of a single randomly chosen sample in each iteration. Therefore, applying a stochastic gradient method for large scale machine learning problems can be computationally efficient \cite{zhang2004solving,alrajeh2014large,bottou2010large,hsieh2008dual}.  In the context of supervised learning, models that are trained by such iterative optimization algorithms are commonly controlled by convergence rate analysis. The convergence rate portrays how fast the optimization error decreases as the number of iterations grows. However, many fast converging algorithms are algorithmically less stable (an algorithm is stable if it is robust to small perturbations in the composition of the learning data set). The stability of an algorithm is considered by measuring the generalization error in terms of the absolute difference between the test and the training error. 

The classical results by Bousquet and Elisseef \cite{bousquet2002stability} showed that a randomized algorithm such as SGM is uniformly stable if for all data sets differing in one element, the learned models produce nearly the same results.  Hardt et al. \cite{hardt2015train} suggested that by choosing reasonable number of iterations SGM generalizes well and prevents overfitting under standard Lipschitz and smoothness assumptions. Therefore, the iterative optimization algorithm can stop long before its convergence to reduce computational cost.  The expected excess risk decomposition has been the main theoretical guideline for this kind of early-stopping criteria.  Motivated by this approach we derived a high probability bound in terms of expected risk for an approximated kernel function considering the stability definition. We proposed that, in the context of supervised learning with a high probability SGM generalizes well for an approximated kernel under proper assumptions.  We showed that with few number of iterations generalization error is independent of model size, and it is a function of number of epochs. In addition, we explored the effect of learning rate choices, and the number of Fourier components on the generalization error.

In this paper, we proved that SGM generalizes well for an approximated kernel under proper assumptions by choosing only few number of iterations while being stable. In particular, we mapped the input data to a randomized low-dimensional feature space to accelerate the training of kernel machines using random Fourier features \cite{rahimi2007random}.  We, then incorporated the approximated kernel function into the primal of SVM to form a linear primal objective function following \cite{chapelle2007training}. Finally, we showed that SGM generalizes well given the approximated algorithm under proper assumptions by incorporating the stability term into the classical convergence bound.

This paper is organized as follows. In section \ref{sec:Pstatement} the detailed problem statement is discussed following the convex optimization setting used for this problem. The theoretical analysis is discussed in section \ref{sec:theory}. This is followed by the numerical results in section \ref{sec:result}. Finally the discussion is provided in section \ref{sec:discussion}.

\section{Preliminaries}\label{sec:Pstatement}

\subsection{Optimization problem}\label{sec:PrimalSVM}
Given a training set $\{(x_i,y_i)\}_{1\leq i \leq n}$, $x_i \in R^d$ , $y_i \in \{+1,-1\}$, a linear hyperplane for SVM problems is defined by $f(x)= w^Tx + b$. Where, $n$ is the number of training examples, and $w$ is the weight coefficient vector. The standard primal SVM optimization problem is shown as: 

\begin{equation}\label{eqn:optSVM1}
\min\limits_{w\in R^d} \frac{\lambda}{2} ||w||^2 + \frac{1}{n} \sum\limits_{i}^n (\max (0,1-y_if(x_i))). 
\end{equation}

Rather than using the original input attributes $x$, we instead used the kernel tricks so that the algorithm would access the data only through the evaluation of $k(x_i, x_j)$. This is a simple way to generate features for algorithms that depend only on the inner product between pairs of input points.  Kernel tricks rely on the observation that any positive definite function $k(x_i, x_j)$ with $x_{i},x_j\in R^d$ defines an inner product and a lifting $\phi$ so that the inner product between lifted data points can be quickly computed as $\langle \phi\left( x\right) ,\phi\left( y\right) \rangle =k(x, y)$. Our goal is to efficiently learn a kernel prediction function $k$ and an associated Reproducing Kernel Hillbert Space $\mathcal H$ as follows:
 
\begin{equation}\label{eqn:optSVM2}
\min\limits_{f\in \mathcal H} \frac{\lambda}{2} ||f||_{\mathcal H}^2 + \frac{1}{n} \sum\limits_{i}^n (\max (0,1-y_if(x_i))).
\end{equation}
Where,
\begin{equation}\label{eqn:fdef}
f(x)=\sum\limits_{i=1}^n \alpha_i K(x_i,x_j),
\end{equation}

However, in large scale problems, dealing with kernels can be computationally expensive. Hence, instead of relying on the implicit lifting provided by the kernel trick, we used explicitly mapping the data to a low-dimensional Euclidean inner product space using a randomized feature map $z :  R^d \rightarrow  R^D$ so that the inner product between a pair of transformed points approximates their kernel evaluation \cite{yang2012nystrom,rahimi2008random}.  Given the random Fourier features, we then learned a linear machine $f(x) = w^Tz(x)$ by solving the following optimization problem:

\begin{equation}\label{eqn:optSVMapx}
\min\limits_{w\in \mathcal R^{2D}} \frac{\lambda}{2} ||w||^2_2 + \frac{1}{n} \sum\limits_{i}^n (\max (0,1-y_iw^Tz(x_i)).
\end{equation}

\subsection{Convex optimization settings}
The goal of our online learning is to achieve minimum expected risk, hence we tried to minimize the loss function. Throughout the paper, we focused on convex, Lipschitz continuous and gradient smooth loss functions, provided their definitions here.

\begin{defn}
A function f is L-Lipschitz continuous if we have $||\nabla f(x)||\leq L$, while implies
\begin{equation}
|f(x)-f(y)|\le L||x-y||, 
\end{equation}
\end{defn}

\begin{defn}
A function f is gradient $\beta$ Lipschitz continuous if we have $||\nabla^2 f(x)||\leq \beta$, while implies
\begin{equation}
|\nabla f(x)-\nabla f(y)|\le \beta||x-y||. 
\end{equation}
\end{defn}

 In the theoretical analysis section we required a convex, Lipschitz continuous and gradient smooth function. Note that a huber-hinge loss function is Lipschitz continuous,  
and it has a Lipschitz continuous gradient which is defined as follows:

$ l_{huber-hinge}(y_i,w^Tz(x_i))=$
  \[
    \left\{
                \begin{array}{ll}
                -4y_iw^Tz(x_i),\;\;\;\;\;\;\;\;\;\;\;\;\;\;\;y_iw^Tz(x_i)<-1\\
               {{(\ 1-y}_iw^Tz(x_i)\ )}^2, \;\; \;\;\ -1 \leq \ y_iw^Tz(x_i) \le 1\\
                 0, \;\;\;\;\;\;\;\;\;\;\;\;\;\;\;\;\;\;\;\;\;\;\;\;\;\;\;\;\;\;\;\;\; y_iw^Tz(x_i)>1\\
                \end{array}
              \right.
  \]
  
Therefore, in this paper, we used the following optimization problem: 

\begin{equation}\label{eqn:risk}
\min\limits_{w\in \mathcal R^{2D}} \sum\limits_{i}^n l_{huber-hinge}(y_i,w^Tz(x_i)).
\end{equation}

For simplicity, the loss function in \eqref{eqn:risk} is denoted by $l(w_t)$. Let $w^*$ be the minimizer of the population risk:

\begin{equation}
\begin{aligned}
R(w)\overset{def}{=} E_{(x,y)}(l(w))
\end{aligned}
\end{equation}

Let $\bar{w}_T=\frac{1}{T}\sum\limits_{t=1}^T w_t$, where $T$ is the maximum iteration for the SGM. According to \cite{lularge} and \cite{nemirovski2009robust}  we have the following Lemma:

\begin{lem}
Let $l(.)$ be a convex loss satisfying $\nabla l(w) \le L$ and let $\eta$ be the constant learning rate. Let $\bar{w}_T=\frac{1}{T}\sum\limits_{t=1}^T w_t$, where $T$ is the maximum SGM iteration. Also, let $w^*$ be the minimizer of the population risk $R(w)= E_{(x,y)}(l(w))$. Then, 
\begin{align}\label{eqn:FODGcon}
R[\bar{w}_T]\le R[w^*] + \frac{||w^*||^2}{2T\eta}+ \frac{\eta}{2}L^2.
\end{align}
\begin{proof}
Note that:
\begin{equation}
\begin{aligned}
||w_{t+1}- w^*||^2= ||w_t - \eta\nabla l_t(w_t)- w^*||^2\\
=||w_t-w^*||^2+ \eta^2||\nabla l_t(w_t)||^2 - 2\eta\nabla l_t(w_t)(w_t-w^*),
\end{aligned}
\end{equation}
and,
\begin{equation}
l_t(w_t) - l_t(w^*) \le \nabla l_t(w_t)(w_t - w^*). 
\end{equation}
Combining these two we have the following: 
\begin{equation}
\begin{aligned}
l_t(w_t) - l_t(w^*) \le \frac{||w_t-w^*||^2-||w_{t+1}- w^*||^2 }{2\eta}\\
+\frac{\eta}{2}||\nabla l_t(w_t)||^2
\end{aligned}
\end{equation}
By summing the above over $T$ and taking average the lemma is proved.
\end{proof}
\end{lem}

From Rahimi \cite{rahimi2007random}, we know that with a high probability of at least $1-2^8(\frac{\sigma_p R}{\epsilon})^2 \exp(\frac{-D\epsilon^2}{4(d+2)})$ there is a
 probability bound for the difference between the approximated kernel value and the exact kernel value. Where $\sigma^2_p=E_p[u^Tu]$ is the second moment of Fourier transform of kernel function. Further the following inequality holds when,  $D=\Omega(\frac{d}{{\epsilon }^2}{\mathrm{log} \frac{{\sigma }_P\ diam(M)}{\epsilon }\ })$:

\begin{equation}
\begin{aligned}
|z(x_i)^Tz(x_j)-k(x_i,x_j)| < \epsilon.
\end{aligned}
\end{equation}

Assuming $k(x_i,x_j)\le 1$ and $z(x_i)^Tz(x_j)\le 1+ \epsilon$, then:

\begin{equation}\label{eqn:FODGw}
||w^*|| \le (1+\epsilon)||f^*||^2_1,
\end{equation}

where $||f^*||_1=\sum\limits_{t=1}^{T}|\alpha^*_t|$, resulting from $f^*(x)=\sum\limits_{t=1}^{T}\alpha^*_tk(x,x_t)$ and $\sum\limits_{t=1}^{T}k(x,x_t)=1$, and  $w^*=\sum\limits_{t=1}^{T}\alpha^*_tz(x_t)$. 
By substituting Equation \eqref{eqn:FODGw} in Equation \eqref{eqn:FODGcon}, with a high probability of $1-2^8(\frac{\sigma_p R}{\epsilon})^2 \exp(\frac{-D\epsilon^2}{4(d+2)})$, we obtain:

\begin{align}\label{eqn:FODGconSec}
R[\bar{w}_T]\le R[w^*] + \frac{(1+\epsilon)||f^*||^2_1}{2T\eta}+ \frac{\eta}{2}L^2 +eL\left\| f^{\ast }\right\| _{1}
\end{align}

Given that, an optimization error is defined as the gap between empirical risk and minimum empirical risk in expectation, and it is denoted by:

\begin{equation}\label{egn:errordef}
\epsilon_{opt}(w)\overset{def}{=}E[R_S[w]-R_S[w^S_*]],  
\end{equation}

where, $S$ denotes a population sample of size $n$ and $R_S[w]$ is the empirical risk defined as:

\begin{equation}
\begin{aligned}
R_S[w]\overset{def}{=}\frac{1}{n}\sum\limits_{i=1}^nl(w;(x_i,y_i))
\end{aligned}
\end{equation}

Note that the expected empirical risk is smaller than the minimum risk , implying:

\begin{equation}\label{eqn:minLessEmp}
E[R_S[w^S_*]]\le R[w^*].
\end{equation}

Hence, based on Equations \eqref{eqn:FODGconSec}, \eqref{egn:errordef} and \eqref{eqn:minLessEmp}, with a high probability of at least $1-2^8(\frac{\sigma_p R}{\epsilon})^2 \exp(\frac{-D\epsilon^2}{4(d+2)})$, we have: 
\begin{equation}
\epsilon_{opt}(w)\le  \frac{(1+\epsilon)||f^*||^2_1}{2T\eta}+ \frac{\eta}{2}L^2. 
\end{equation}

\begin{lem}
Let l be a convex loss function that is Lipschitz continuous and $\nabla(l(w))\le L$. Let $||f^*||_1=\sum\limits_{t=1}^{T}|\alpha^*_t|$; resulting from $f^*(x)=\sum\limits_{t=1}^{T}\alpha^*_tk(x,x_t)$ and $\sum\limits_{t=1}^{T}k(x,x_t)=1$. Also let $w^*=\sum\limits_{t=1}^{T}\alpha^*_tz(x_t)$.
 Suppose we make a single pass SGM over all the samples $(T=n)$, and by choosing $\eta=\frac{||f^*||_1}{L\sqrt[]{n}}$, then with a high probability $1-2^8(\frac{\sigma_p R}{\epsilon})^2 \exp(\frac{-D\epsilon^2}{4(d+2)})$, the classical convergence bound in \eqref{eqn:FODGconSec} becomes:
 
\begin{equation}\label{eqn:singlepass}
E[R[\bar{w}_n]]\le R[w_*]+ \frac{(2+\epsilon)||f^*||_1 L}{(2)\sqrt[]{n}}. 
\end{equation}
\end{lem}

Knowing that,

\begin{equation}\label{eqn:errorstab}
E[R[w]]\le E[R_S[w^S_*]]+ \epsilon_{opt}(w)+\epsilon_{stab},
\end{equation}

Where $\epsilon_{stab}$ is the stability error satisfying $\epsilon_{stab} \le \frac{TL^2\eta}{n}$, and given that the function is $L$-Lipschitz continuous and $\beta$-smooth. We know that $\epsilon_{opt}$ will decrease with the number of SGM iterations while $\epsilon_{stab}$ increases. Hardt et. al. in \cite{hardt2015train} showed that given few number of iterations and by balancing $\epsilon_{stab}$ and $\epsilon_{opt}$, the generalization error will decrease.  In the next section, we explored to see whether using SGM for an approximated algorithm which favors in terms of computational cost would generalize well by choosing few number of iterations while being stable. 

\section{Generalization of SGM for an approximated algorithm}\label{sec:theory}
\begin{thm}
Let l be $L$-Lipschitz continuous and $\beta$-smooth. $w^S_*$ is the minimizer of the empirical risk and $R_S[w]=\frac{1}{n}\sum\limits_{i=1}^{n}l(w^T_tz(x_t);y_t)$. Let $||f^*||_1=\sum\limits_{t=1}^{T}|\alpha^*_t|$,  where $f (x)=\sum\limits_{t=1}^{T}\alpha^*_tk(x,x_t)$ and $\alpha^*_t$ is the coefficient of the $i$th support vector. For the maximum iteration $T$ of the SGM, with high probability of $1-2^8(\frac{\sigma_p R}{\epsilon})^2 \exp(\frac{-D\epsilon^2}{4(d+2)})$, we have:
\begin{equation}\label{eqn:boundstab}
E[R[\bar{w}_T]]\le E[R_S[w^S_*]] + \frac{||f^*||_1L\sqrt{1+\epsilon}}{\sqrt{n}}\sqrt{\frac{n+2T}{T}}.
\end{equation}
\begin{proof}
Recall that with a high probability, $\epsilon_{opt}(\bar{w}_T)\le  \frac{(1+\epsilon)||f^*||^2_1}{2T\eta}+ \frac{\eta}{2}L^2$. Also recall that $\epsilon_{stab}\le \frac{TL^2\eta}{n}$. Then by substituting these two terms in \eqref{eqn:errorstab}, for every $f^*$, with a high probability $1-2^8(\frac{\sigma_p R}{\epsilon})^ \text{exp}\frac{-D\epsilon^2}{4(d+2)}$, we have:
\begin{equation}\label{eqn:finth}
E[R[\bar{w}_T]]- E[R_S[w^S_*]]\le  \frac{(1+\epsilon)||f^*||^2_1}{2T\eta}+ \frac{\eta}{2}L^2 + \frac{TL^2\eta}{n}.
\end{equation}
By taking the gradient of the right hand side of \eqref{eqn:finth} with respect to $\eta$, the optimal $\eta$ is:
\begin{equation}\label{eqn:eta2}
\eta=\frac{||f^*||_1\sqrt[]{(1+\epsilon)n}}{L\sqrt[]{T(n+2T)}}.
\end{equation}
By substituting the optimal $\eta$ in Equation \eqref{eqn:finth} the theorem is proved. 
\end{proof}
\end{thm}

The above theorem suggests that with a high probability SGM generalizes well for an approximated kernel for $L$-Lipschitz continuous and $\beta$-smooth loss function. In general, the optimization error ($\epsilon_{opt}$) decreases with the number of SGM iterations while the stability ($\epsilon_{stab}$) increases. From \eqref{eqn:boundstab} we can claim that as the number of iteration increases, $\epsilon_{opt}$ and $\epsilon_{stab}$ will become less balanced. Thus, choosing few number of iterations would balance $\epsilon_{opt}$ and $\epsilon_{stab}$ suggesting a stable SGM.

By setting the $\eta$ to \eqref{eqn:eta2} when $T = n$  the generalization error bound for an approximated kernel based on random Fourier features is given by,
\begin{equation}\label{eqn:computationCost}
E[R[\bar{w}_T]]\le E[R_S[w^S_*]] \leq O(\frac{1}{\sqrt{n}})
\end{equation}

Our generalization bound has a convergence rate of $O(\frac{1}{\sqrt{n}})$, where compared with the rate achieved by \cite{shalev2011pegasos} of $O(\frac{1}{n})$
is significantly more efficient. Recall that from \cite{rahimi2007random} number of random Fourier components is given by $D=\Omega(\frac{d}{{\epsilon }^2}{\mathrm{log} \frac{{\sigma }_P\ diam(M)}{\epsilon }\ })$. By setting $\epsilon=\frac{1}{\sqrt[]{n}}$ we require to sample $D = O(n)$ Fourier features in order to achieve a high probability. A regular classifier $f(x)=\sum\limits_{i=1}^N \alpha_i K(x_i,x)$, requires $O(nd)$ time to compute; however, with the randomized feature maps $f(x)= w^Tz(x)$ only $O(D)$ operations is required. Thus, using reasonable number of iterations an approximated kernel learning machine is faster than a regular kernel method with an advantage of preventing overfitting, and making it more practical for large-scale kernel learning.

\section{Experimental results}\label{sec:result}
Theatrically we proved that an approximated Fourier primal support vector machine is stable providing a smooth loss function and relatively sufficient number of steps. Thus, given reasonable number of epochs stochastic gradient method would generalize well, and prevent possible overfitting. We numerically showed the effect of three parameters; model size, number of Fourier components and learning rate choices on the stability. Table \eqref{tab:1} shows the description of four binary classification datasets used for the analysis. These datasets can be downloaded from UCI machine learning repository website.

\begin{table}
\caption{Statistics of binary classification datasets}\label{tab:1}      
\begin{tabular}{llll}
\hline\noalign{\smallskip}
Dataset & Sample size & Dimension \\
\noalign{\smallskip}\hline\noalign{\smallskip}
spambase & 4601 & 57 \\
german & 1000 & 24 \\
svmguide3 & 1284 & 21\\
Pima Indians Diabetes & 768 & 8\\
\noalign{\smallskip}\hline
\end{tabular}
\end{table}

\begin{figure*}
\centering
\includegraphics[width=6in, height=1.7in, keepaspectratio=false]{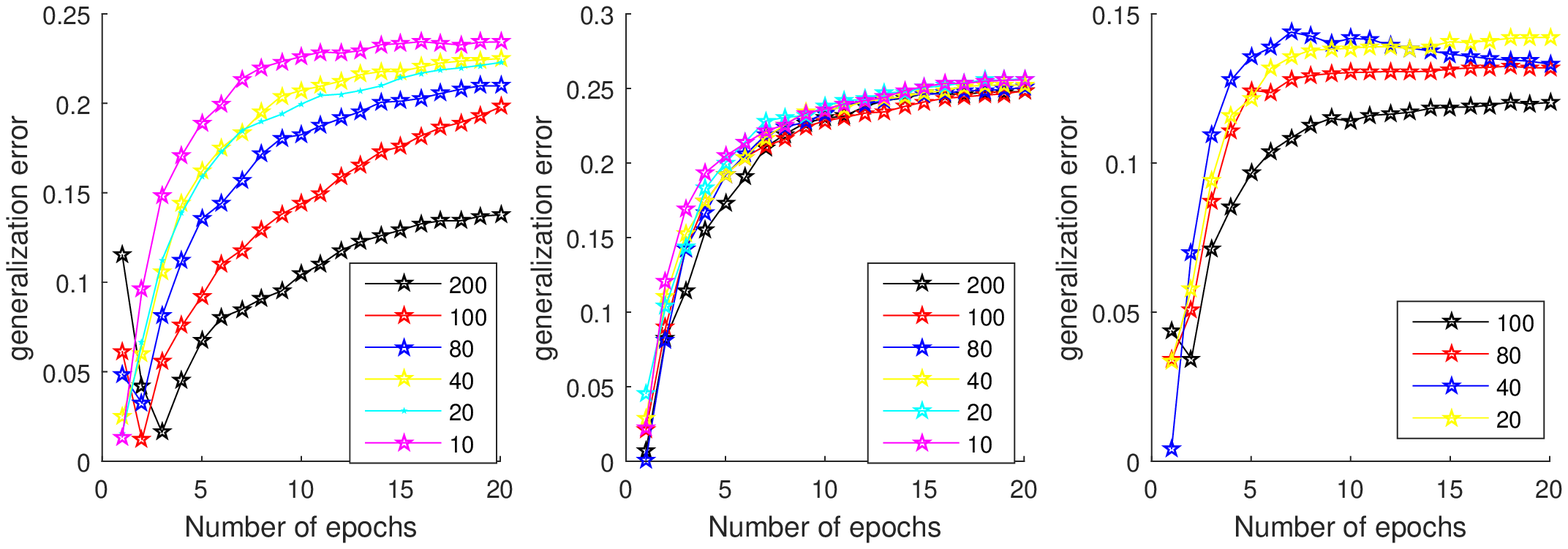}
\caption{ the effect of random Fourier features on the generalization error for spambase, german and svmguide3 data sets}\label{fig:2} 
\end{figure*}

\begin{figure} 
\includegraphics[width=3.8in, height=3.2in, keepaspectratio=false]{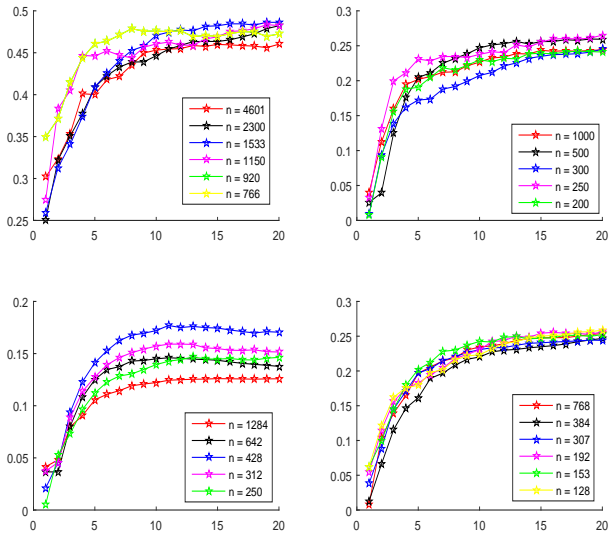}  
\caption{The effect of model size on the generalization error for spambase, german and svmguide3 and Pima Indians Diabetes datasets}\label{fig:1}
\end{figure}

\begin{figure} 
\includegraphics[width=3.5in, height=3.2in, keepaspectratio=false]{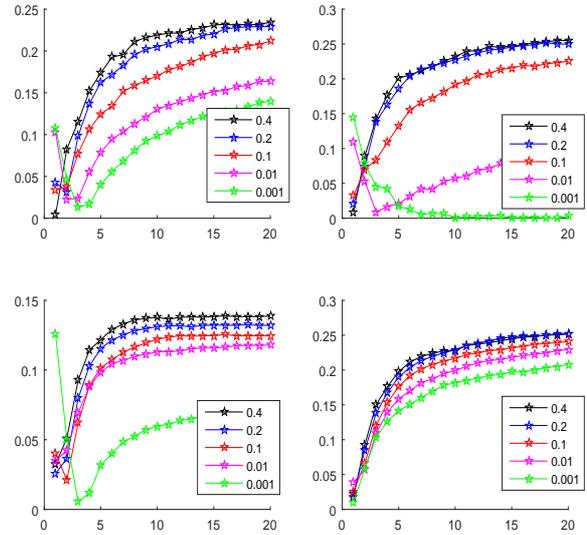}
\caption{ Generalization error for varying learning rates for spambase, german and svmguide3 and Pima Indians Diabetes datasets}\label{fig:3}
\end{figure}

In Figure \ref{fig:2}, we showed the effect of the number of random Fourier features on the generalization error. The result showed that the generalization error is a function of number of random Fourier features. The approximated kernel performs nearly the same as the exact kernel based learning by choosing large number of Fourier components. This means if we sample more number of Fourier components, the approximation of kernel function is more accurate. In general, increasing number of Fourier components leads to a better approximation and thus a lower testing error \cite{sutherland2015error}. On the other hand, the computation cost is proportional to the number of Fourier components. Hence, we performed a simulation based experiment for each data set to find the best number of Fourier components. For the computation cost purposes we restricted the maximum number of Fourier components to a maximum of 200 features. In addition, we used numerical examples to demonstrate the dependence of the generalization error on the number of epochs and its independence on the sample size. Figure \ref{fig:1}, shows the generalization error for different epochs. We defined epochs as the number of complete passes through the training set. The results demonstrated that the generalization error is a function of number of epochs and not the model size.  Choosing a proper learning rate for stochastic gradient method is crucial. Figure \ref{fig:3}, shows the strong impact of learning rate choices on the generalization error. We conducted an experiment for searching the best learning rate for all data sets.

\section{Discussion}\label{sec:discussion}
In this paper we measured the stability of stochastic gradient method (SGM) for learning an approximated Fourier primal support vector machine. We demonstrated that a large-scale approximated online kernel machine using SGM is stable with a high probability. The empirical results showed that the generalization error is a function of number of epochs and independent of model size. We also showed the strong impact of learning rate choices and number of Fourier components. Moreover, in this paper we utilized SGM to solve an approximated primal SVM.  Utilizing random Fourier features induced variance, which slowed down the convergence rate.  One way to tackle this problem is using variance reduction methods such as stochastic variance reduced gradient (SVRG) \cite{johnson2013accelerating}.

\section{Conflict of Interest}
The authors declare that they have no conflict of interest.

 \begin{acknowledgements}
Authors are grateful for the tremendous support and help of Professor Maryam Fazel
 \end{acknowledgements}

\bibliographystyle{spmpsci}      % mathematics and physical sciences
\bibliography{refEE546}   % name your BibTeX data base

\end{document}